\documentclass[a4paper,reqno,12pt]{amsart}
\usepackage{amssymb,euscript,bbold}
\usepackage{array}
\DeclareMathOperator{\vol}{\mathrm{vol}}
\DeclareMathOperator{\tr}{tr}
\newcommand{\RR}{\mathbb{R}}

\newcommand{\U}{\mathrm{U}}
\newcommand{\SU}{\mathrm{SU}}
\renewcommand{\Re}{\mathrm{Re}}
\renewcommand{\Im}{\mathrm{Im}}
\renewcommand{\Sp}{\mathrm{Sp}}
\newcommand{\Spin}{\mathrm{Spin}}

\newcommand{\CC}{\mathbb{C}}
\newcommand{\EE}{\mathbb{E}}
\newcommand{\MM}{\mathbb{M}}
\newcommand{\HH}{\mathbb{H}}
\newcommand{\SO}{\mathrm{SO}}
\newcommand{\ZZ}{\mathbb{Z}}
\newcommand{\Cl}{\mathrm{C}\ell}

\newcommand{\half}{\tfrac{1}{2}}
\newcommand{\fS}{\mathfrak{S}}

\newcommand{\M}{\mathsf{M}}
\newcommand{\eM}{\EuScript{M}}
\newcommand{\D}{\mathsf{D}}
\newcommand{\1}{\mathbb{1}}
\newcommand{\AdS}{\mathrm{AdS}}
\newtheorem{thm}{Theorem}

\begin{document}

\title[branes at angles and calibrated geometry]{Branes at
angles and calibrated geometry}
\author[acharya]{BS Acharya}
\email{r.acharya@qmw.ac.uk}
\author[figueroa-o'farrill]{JM Figueroa-O'Farrill}
\email{j.m.figueroa@qmw.ac.uk}
\author[spence]{B Spence}
\email{b.spence@qmw.ac.uk}
\address[]{\begin{flushright}Department of Physics\\
Queen Mary and Westfield College\\
Mile End Road\\
London E1 4NS, UK\end{flushright}}
\begin{abstract}
In a recent paper, Ohta \& Townsend studied the conditions which must
be satisfied for a configuration of two intersecting $\M5$-branes at
angles to be supersymmetric.  In this paper we extend this result to
any number of $\M5$-branes or any number of $\M2$-branes.  This is
accomplished by interpreting their results in terms of calibrated
geometry, which is of independent interest.
\end{abstract}
\maketitle

\section{Introduction}

One of the important lessons to be drawn from the recent developments
in nonperturbative superstring theory is the fact that this is a
theory of more than strings.  Branes have been shown to play a
decisive role and understandably much of the recent effort in
superstring theory has shifted towards the study of branes.  Branes
are fascinating objects and understanding their dynamics and their
geometry is one of the more challenging problems facing superstring
theory.

In the absence of any other background field beside the metric, a
$p$-brane is described by the Dirac--Nambu--Goto action, which has as
classical solutions minimal immersions of the world-volume of the
$p$-brane into the spacetime.  Minimal means that the mean curvature
vector of the immersion has zero trace.  In the case of static branes
or of euclidean branes, the volume which is being extremised is the
euclidean volume.  One way to generate minimal submanifolds (in fact,
homologically volume-minimising) is the method of calibrations
introduced by Harvey \& Lawson \cite{HarveyLawson}.  In practice we
are not interested in branes which merely minimise volume, but in
those configurations which preserve some of the supersymmetry of the
underlying theory.  These supersymmetric configurations are
volume-minimising, but the converse is not true.  Therefore
supersymmetric brane configurations represent a refinement of the
notion of minimal submanifold, but remarkably one for which the theory
of calibrations is well suited.  Even more unexpectedly, calibrations
seem to play a role in the study of branes in the presence of other
background fields, for example the $B$-field.  In this case the brane
dynamics is described by actions of the Dirac--Born--Infeld type,
whose classical solutions are no longer minimal.  Nevertheless in some
cases one can still describe these solutions locally using slight
generalisations of calibrations, as in the recent work of Stanciu on
$\D$-branes on coset models \cite{Sonia}.

The purpose of this paper is to place the study of the intersecting
branes in the context of calibrated geometry and to apply these
methods to the supersymmetric configurations of intersecting
$\M$-branes.  We shall focus primarily on $\M5$-branes.  $\M$-theory
has eleven-dimensional supergravity as its low-energy limit, whose
bosonic fields are the metric of an eleven-dimensional lorentzian spin
manifold and a closed 4-form.  The $\M5$-brane is a solution to the
classical equations of motion where the metric takes the form of a
warped product of six-dimensional Minkowski spacetime with a
five-dimensional conformally euclidean space.  The six-dimensional
spacetime is to be interpreted as the worldvolume of a
five-dimensional extended object---the $\M5$-brane---and the
eleven-dimensional metric is to be thought of as describing the
exterior spacetime around such an object, not unlike the Schwarzschild
black hole solution.  Although the $\M5$-brane preserves only one half
of the supersymmetry present in eleven-dimensional supergravity, it
nevertheless interpolates between two solutions with maximal
supersymmetry: eleven-dimensional Minkowski spacetime asymptotically
far away from the brane, and $\AdS_7 \times S^4$ near the brane
horizon.  There are a wealth of other solutions known to
eleven-dimensional supergravity, among them the $\M$-wave and the
$\M2$-brane, both preserving one half of the supersymmetry, as well as
solutions preserving a smaller fraction of the supersymmetry which can
be interpreted as intersecting branes.  For a recent review on
intersecting branes and a guide to extensive literature, see
\cite{Jerome}.

The complete classification of supersymmetric solutions of
eleven-dimensional supergravity would be an important step towards
understanding the true nature of $\M$-theory; but this seems to be a
very difficult problem.  A more manageable task seems to be the
classification of supersymmetric configurations which locally look like
intersecting $\M$-branes.  Townsend \cite{Townsend-Mbranes} initiated
the classification of the supersymmetric configurations of a pair of
intersecting $\M5$-branes, a classification completed with Ohta in
\cite{OhtaTownsend}.  Their method can be summarised as follows.  One
starts from a configuration of two parallel $\M5$-branes, and rotates
one of the two branes.  The rotation matrix depends roughly on a set
of five characterising angles.  The angles for which the configuration
preserves some supersymmetry are then determined via an explicit
computation.  This method is not practicable for configurations
involving more than two branes, so it seems that a more indirect
approach is needed in order to tackle the general case.  This is the
purpose of the present paper.  We will rederive their results and, at
no extra cost, extend them to an arbitrary number of intersecting
$\M5$-branes.  The method also works for configurations of
$\M2$-branes and we briefly study them as well.

This paper is organised as follows.  We start in Section 2 by
rederiving the results of Ohta \& Townsend \cite{OhtaTownsend} in a
more invariant fashion.  Because the notation natural to our method
differs slightly from what is standard in the brane literature, this
section also serves as a dictionary.  In Section 3 we briefly
introduce the basic notation and concepts of calibrated geometry.  We
discuss grassmannians, volume minimisation, calibrations on manifolds,
and its relation with spinors and special holonomy.  In Section 4 we
discuss the Angle Theorem and the Nance calibrations, which answer the
question of when do two intersecting planes minimise volume.  In
Section 5 we specialise these constructions to the case of
intersecting branes.  We first interpret the results of Section 2 in
terms of calibrations and then extend them to include any number of
intersecting $\M5$-branes at angles.  We also treat the case of
$\M2$-branes.  Finally in Section 6 we offer some conclusions.

After this work had been completed, two papers appeared
\cite{GibbonsPapadopoulos,GLW} which also address the problem of
intersecting branes from the point of view of calibrated geometry.
Except for the fact that calibrated geometry plays a crucial role in
all three papers, there is little substantial overlap between this
paper and the other two.  Earlier papers on branes which also make
contact with calibrated geometry are \cite{GT,BBS,BSV,OOY,BBMOOY}.

\section{Supersymmetric pairs of $\M5$-branes at angles}

In this section we will rederive some of the results of Ohta \&
Townsend \cite{OhtaTownsend} and we will also set out our notation.
Let us consider the $\M5$-brane solution.  Let $(x^\mu)$ denote the
eleven-dimensional coordinates, where $(x^0,x^1,\ldots,x^5)$ are
coordinates along the brane and $(x^6,\ldots,x^9,x^{\natural})$ are
coordinates transverse to the brane.  Far away from the brane, the
metric is asymptotically flat, so that the Killing spinors of the
supergravity solution have constant asymptotic values $\varepsilon$,
obeying
\begin{equation}\label{eq:halfsusy}
\Gamma_{012345} \varepsilon = \varepsilon~,
\end{equation}
where $\varepsilon$ is a real $32$-component spinor of
$\Spin_0(10,1)$ contained in the Clifford algebra $\Cl(10,1)$
generated by the $\Gamma_M$.
Provided we only deal with one brane, it is possible to choose
coordinates so that the brane is stretched along these directions; but
the moment we have to consider two or more branes, particularly if
they intersect non-orthogonally, this notation becomes cumbersome,
since not all branes can be described so conveniently.  Moreover our
aim in this paper is not to analyse the global properties of branes,
but only their local properties at the point of intersection.  In
fact, we could be analysing singularities in a single brane which is
immersed (rather than embedded) in the spacetime.  We will therefore
recast the work of \cite{OhtaTownsend} in terms of tangent planes at a
point to the branes themselves.

So we fix a point $x$ in the spacetime $M$ and we fix an orthonormal
frame for the tangent space $e_0,e_1,\ldots,e_9,e_\natural$, which
allows us to identify the tangent space $T_xM$ with eleven-dimensional
Minkowski spacetime $\MM^{10,1}$.  We will further decompose
$\MM^{10,1} = \EE^{10} \oplus \RR e_0$.  This decomposition is
preserved by an $\SO(10)$ subgroup of $\SO(10,1)$.  As in
\cite{OhtaTownsend} we will restrict ourselves to configurations for
which the tangent plane to the worldvolume of a given $\M5$-brane
passing through $x$ is spanned by $e_0, v_1, \ldots, v_5$, where $v_i$
are orthonormal vectors in $\EE^{10}$.  Suppose moreover that the
brane is given the orientation defined by $e_0\wedge v_1 \wedge \cdots
\wedge v_5$.  We will therefore be able to associate with each such
brane at $x$ a 5-vector $\xi = v_1 \wedge \cdots \wedge v_5$ in
$\bigwedge^5\EE^{10}$.  Conversely, to any given unit simple 5-vector
$\xi = v_1 \wedge \cdots \wedge v_5$, we associate an oriented 5-plane
given by the span of the $v_i$.  The condition for supersymmetry
\eqref{eq:halfsusy} can be rewritten more generally as
\begin{equation}\label{eq:fundsusy}
(e_0 \wedge \xi ) \cdot \varepsilon = \varepsilon~,
\end{equation}
where $\cdot$ stands for Clifford multiplication and where we have
used implicitly the isomorphism of the Clifford algebra
$\Cl(10,1)$ with the exterior algebra $\bigwedge\MM^{10,1}$.  When
$\xi = e_1\wedge e_2 \wedge \cdots\wedge e_5$, equation
\eqref{eq:fundsusy} agrees with equation \eqref{eq:halfsusy}.

Now suppose that we are given two $\M5$-branes through $x$ with
tangent planes $\xi$ and $\eta$.  This configuration will be
supersymmetric if there exists a nonzero spinor $\varepsilon$ for
which
\begin{equation*}
(e_0 \wedge \xi ) \cdot \varepsilon = \varepsilon
\qquad\text{and}\qquad
(e_0 \wedge \eta ) \cdot \varepsilon = \varepsilon~.
\end{equation*}
Because $\SO(10)$ acts transitively on the space of $5$-planes, there
exists a rotation $R$ in $\SO(10)$ which transforms $\xi$ to $\eta$.
Because $R$ is conjugate to any given maximal torus of $\SO(10)$,
there exists a choice of orthonormal frame $e_i$ for which $\xi =
e_1\wedge e_3 \wedge \cdots \wedge e_9$ and
\begin{equation*}
\eta = R(\theta)\xi = (\cos\theta_1 e_1 + \sin\theta_1 e_2) \wedge
\cdots \wedge (\cos\theta_5 e_9 + \sin\theta_5 e_\natural)~,
\end{equation*}
where $R(\theta)$ is the block-diagonal matrix
\begin{equation*}
R(\theta) = 
\begin{pmatrix}
R_{12}(\theta_1) & & & & \\
& R_{34}(\theta_2) & & & \\
& & R_{56}(\theta_3) & & \\
& & & R_{78}(\theta_4) & \\
& & & & R_{9\natural}(\theta_5)
\end{pmatrix}~,
\end{equation*}
each $R_{jk}(\vartheta)$ being the rotation by an angle $\vartheta$ in
the 2-plane spanned by $e_j$ and $e_k$.  The angles $(\theta_i)$ are
of course not unique, since even after having conjugated $R$ to a
given maximal torus, we can still act with Weyl transformations.  As
we will discuss below, all this means is that we can choose an
ordering for the angles and certain signs.  Let $\eM$ denote the space
of possible angles; that is, the quotient of the maximal torus of
$\SO(10)$ by the action of the Weyl group.  If $\theta \equiv
(\theta_i)$ are any five angles, we will let $[\theta] \in \eM$ denote
their equivalence class under the action of the Weyl group.  The
subset $\eM_{\mathrm{susy}} \subset \eM$ consists of those angles
$[\theta]$ for which the intersecting brane configuration defined by
$\xi$ and $R(\theta)\xi$ preserves some supersymmetry.  In other
words, $\eM_{\mathrm{susy}}$ is the subset of $\eM$ for which there is
at least one nonzero spinor $\varepsilon$ which solves the following
equations:
\begin{equation}\label{eq:OT}
(e_0 \wedge \xi ) \cdot \varepsilon = \varepsilon
\qquad\text{and}\qquad
(e_0 \wedge R(\theta)\xi ) \cdot \varepsilon = \varepsilon~.
\end{equation}
For each point $[\theta]$ in $\eM$, let $32\nu([\theta])$ be equal to
the number of linearly independent solutions $\varepsilon$ to
\eqref{eq:OT}.  Therefore $\nu$ defines a (discontinuous) function on
$\eM$ taking the values $0, \frac{1}{32}, \frac{1}{16}, \frac{3}{32},
\ldots, \half$, corresponding to the fraction of the supersymmetry
preserved by the configuration.  In particular, $\nu([\theta])=\half$
if and only if all the angles vanish, whereas $\nu([\theta])\neq 0$ if
and only if $[\theta]$ belongs to $\eM_{\mathrm{susy}}$.

Let $\widehat R$ denote any one of the two possible lifts to
$\Spin(10)$ of the $\SO(10)$ rotation $R$.  Then the second equation
in \eqref{eq:OT} can be written as follows:
\begin{equation*}
\widehat{R}(\theta)\cdot (e_0 \wedge \xi) \cdot
\widehat{R}(\theta)^{-1} \cdot \varepsilon = \varepsilon~.
\end{equation*}
Using the fact that
\begin{equation*}
(e_0 \wedge \xi) \cdot \widehat{R}(\theta)^{-1} =
\widehat{R}(\theta)\cdot (e_0 \wedge \xi)~,
\end{equation*}
together with the first equation in \eqref{eq:OT}, we arrive at
\begin{equation}\label{eq:OT2}
\widehat{R}(\theta)^2 \cdot \varepsilon = \varepsilon~,
\end{equation}
with the same equation resulting for the other possible lift
$-\widehat{R}(\theta)$.
Notice that $\widehat{R}(\theta)^2$ is given explicitly by
\begin{equation*}
\widehat{R}(\theta)^2 = (\cos\theta_1 - \sin\theta_1 \Gamma_{12})\cdot
\cdots \cdot (\cos\theta_5 - \sin\theta_5 \Gamma_{9\natural})
\in\Cl(10)~,
\end{equation*}
which is an element in the maximal torus of $\Spin(10)$ corresponding
to the chosen maximal torus for $\SO(10)$.  Now $\Spin(10)$ has two
complex half-spin representations $\Delta_\pm$, satisfying $\Delta_-^*
\cong \Delta_+$.  Therefore their direct sum $\Delta_- \oplus
\Delta_+$ has a real structure.  The underlying real representation
$\Delta$, defined by $\Delta\otimes_\RR \CC = \Delta_- \oplus
\Delta_+$, is the real spinor representation of $\Spin(10,1)$ to which
$\varepsilon$ belongs.  Under $\Spin(10)$ it is convenient to think of
$\varepsilon$ as a conjugate pair of spinors, $\varepsilon =
(\psi,\psi^*) \in \Delta_-\oplus\Delta_+$, and \eqref{eq:OT2} then
becomes the statement that $\psi \in \Delta_-$ is invariant under the
action of $\widehat{R}(\theta)^2 \in \Spin(10)$.  The real and
imaginary parts of each such $\psi$ give two real solutions of
\eqref{eq:OT2}, but exactly one out of each such pair also obeys the
first equation in \eqref{eq:OT}.  All the weights of the half-spin
representation $\Delta_-$ belong to the Weyl orbit of the highest
weight: $\half(1,1,1,1,-1)$ in the chosen basis.  Therefore, up to a
Weyl transformation, it is enough to evaluate $\widehat{R}(\theta)^2$
on the highest weight vector of $\Delta_-$, yielding $\exp
i\left(\theta_1 + \theta_2 + \theta_3 + \theta_4 - \theta_5\right)$.
Therefore we arrive at the following elegant characterisation of
$\eM_{\mathrm{susy}}$ \cite{OhtaTownsend}:
\begin{equation}\label{eq:msusy}
\eM_{\mathrm{susy}} = \left\{[\theta]\in\eM \left| \sum_{i=1}^4
\theta_i \equiv \theta_5 \mod 2\pi\right.\right\}~.
\end{equation}
The other weights in $\Delta_-$ are obtained from $\half(1,1,1,1,-1)$
by changing the signs of any pair(s) of entries.  Given
$[\theta]\in\eM_{\mathrm{susy}}$, then $32\nu([\theta])$ is equal to
the number of solutions of $\sum_i \sigma_i \theta_i \equiv 0 \mod
2\pi$, where $(\sigma_i)\in \ZZ_2^5$ are signs such that their product
is $-1$.

The foregoing analysis, however, does not seem very practical when one
is considering more than two intersecting 5-planes.  For example,
consider three 5-planes $\xi$, $\eta_1$ and $\eta_2$.  There are
rotations $R_1$ and $R_2$ which transform $\xi$ into $\eta_1$ and
$\eta_2$ respectively; but they will belong in general to different
maximal tori.  Therefore it will not be possible to describe them both
in terms of angles.  This makes the analysis of the analogous equation
to \eqref{eq:OT2} much more involved.  Our approach to this problem
(which appears in Section 5) will involve a re-interpretation of
\eqref{eq:msusy} in the language of calibrated geometry, a topic to
which we now turn.

\section{Calibrated geometry}

In this section we describe the method of calibrations and its
relation with holonomy and spinors.  The foundations of the theory are
to be found in the beautiful paper of Harvey \& Lawson
\cite{HarveyLawson}, and a summary of some of the basic theory is
the expository article \cite{Morgan-monthly} by Morgan.  As mentioned
in the Introduction, calibrations are useful in constructing
globally minimal submanifolds of riemannian manifolds, especially
those with special holonomy.  For our present purposes the crucial
property of calibrations is their relation with spinors.  A leisurely
account of this aspect of the theory can be found in Harvey's book
\cite{Harvey}.

\subsection{Calibrations and volume minimisation}

We will let $G(p,n)$ denote the grassmannian of oriented $p$-planes in
the euclidean space $\EE^n$.  It can naturally be identified with a
subset of the unit sphere in $\RR^{\binom{n}{p}}$, whence it is
compact.  Indeed, given an oriented $p$-plane, let
$e_1,e_2,\ldots,e_p$ be an oriented orthonormal basis and consider the
$p$-vector $\xi = e_1 \wedge e_2 \wedge \cdots \wedge e_p \in
\bigwedge^p \EE^n$.  The norm of any simple $p$-vector $v_1 \wedge v_2
\wedge \cdots \wedge v_p$ is given by
\begin{equation*}
\| v_1 \wedge v_2 \wedge \cdots \wedge v_p \| = | \det\langle
v_i,v_j\rangle|~,
\end{equation*}
from where it follows that $\xi$ has unit norm in $\bigwedge^p \RR^n
\cong \RR^{\binom{n}{p}}$.  Conversely, every unit simple $p$-vector
$\xi = e_1 \wedge e_2 \wedge \cdots \wedge e_p \in \bigwedge^p \EE^n$
defines an oriented $p$-plane with basis $e_1,e_2,\ldots,e_p$.  In
what follows we will make no distinction between a unit simple
$p$-vector and the associated oriented $p$-plane.

Now let $\varphi \in \bigwedge^p \left(\EE^n\right)^*$ be a (constant
coefficient) $p$-form on $\EE^n$.  It defines a linear function on
$\bigwedge^p \EE^n$, which restricts to a continuous function on the
grassmannian $G(p,n)$.  Because $G(p,n)$ is compact, this function
attains a maximum, called the {\em comass\/} of $\varphi$ and denoted
$\|\varphi\|^*$.  Computing the comass of a $p$-form is a difficult
problem which has not been solved but for the simplest of forms
$\varphi$, or for those forms which can be built out of spinors.  If
$\varphi$ is normalised so that it has unit comass $\|\varphi\|^* =
1$, then it is called a {\em calibration\/}.  Let $G(\varphi)$ denote
those points in $G(p,n)$ on which $\varphi$ attains its maximum.
$G(\varphi)$ is known as the {\em $\varphi$-grassmannian\/}.  The
subset $\cup_\varphi G(\varphi) \subset G(p,n)$, where the union runs
over all calibrations $\varphi$, defines the {\em faces of
$G(p,n)$\/}.  The name comes from the fact that if we think of
$G(p,n)$ as a subset of the vector space $\RR^{\binom{n}{p}}$, then
$G(\varphi)$ is the contact set of $G(p,n)$ with the hyperplane
$\varphi(\xi)=1$ on $\RR^{\binom{n}{p}}$.  Now, because $\varphi$ is a
calibration, $\varphi(\xi) \leq 1$ and hence $G(p,n)$ lies to one side
of that hyperplane.

A $p$-submanifold $N$ of $\EE^n$, all of whose tangent
planes belong to $G(\varphi)$ for a fixed calibration $\varphi$, has
minimum volume among the set of all submanifolds $N'$ with the same
boundary.  This is because
\begin{equation*}
\vol N = \int_N \varphi = \int_{N'} \varphi \leq \vol N'~,
\end{equation*}
where the second equality follows by Stokes' theorem.  This is a
generalisation of the notion of a geodesic.  Indeed, the grassmannian
of oriented lines $G(1,n)$ is just the unit sphere
$S^{n-1}\subset\RR^n$, whose faces are obviously points.  Hence the
tangent spaces of a one-dimensional submanifold $L$ belong to the same
face if and only if $L$ is a straight line.  Notice that there is a
duality between $p$-dimensional and $p$-codimensional submanifolds; in
fact, if $\varphi$ is a calibration so is $\star\varphi$.  Hence
hyperplanes in $\EE^n$ are also (locally) volume-minimising.

This theory is not restricted to constant coefficient calibrations in
$\EE^n$.  In fact, we can work with $d$-closed forms $\varphi$ in any
riemannian manifold $(M,g)$.  The comass of $\varphi$ is now the
supremum (over the points in $M$) of the comasses at each point.  If
$M$ is compact, this supremum exists.  A calibration is now a
$d$-closed form normalised to have unit comass; or equivalent one
which satisfies
\begin{equation*}
\varphi_x(\xi) \leq \vol \xi \qquad\text{for all oriented tangent
$p$-planes $\xi$ at $x$.}
\end{equation*}
Notice that there may be points in $M$ for which the
$\varphi$-grassmannian is empty.  The same argument as before shows
that if $N\subset M$ is a submanifold for which $\varphi$ coincides
with the volume form, then $N$ is homologically volume-minimising.
Such manifolds are called {\em calibrated\/}.  Of course, this
crucially necessitates that $\varphi$ be $d$-closed.  Remarkably,
there are physically interesting situations where a submanifold
(indeed, a $\D$-brane) is `calibrated' by a form $\varphi$ of unit
comass, but which is {\em not\/} $d$-closed \cite{Sonia}.

\subsection{Calibrations and holonomy}

As we mentioned above, although every closed $p$-form (at least on a
compact manifold) can be normalised to be a calibration, the
computation of the comass has proven to be a very difficult problem
even for constant calibrations in $\EE^n$.  Luckily, on a riemannian
manifold of reduced holonomy, the comass of a parallel form is
relatively straightforward to compute.  This was well-known for the
case of K\"ahler manifolds, and to a lesser extent for quaternionic
K\"ahler manifolds \cite{Berger}, by the time Harvey \& Lawson wrote
their foundational essay on calibrated geometry.  In this paper, they
discovered a rich geometry of calibrated submanifolds on Ricci-flat
manifolds with $\SU(n)$, $G_2$, and $\Spin(7)$ holonomy.  These
(together with hyperk\"ahler manifolds) are precisely the manifolds
admitting parallel spinors (see, for example, \cite{LM,Wang}) and
hence the cases that play a role in our approach to supersymmetric
brane configurations.

The classic example is K\"ahler geometry.  Let $M$ be a
$2n$-dimensional K\"ahler manifold---that is, a manifold of $\U(n)$
holonomy.  Then the normalised powers $\omega^p/p!$ of the K\"ahler
form are calibrations and its calibrated submanifolds are precisely
the complex submanifolds.  This was first proven by Federer as a
consequence of Wirtinger's inequality.  A local model for this
calibrated geometry is given by considering $\CC^n = \RR^{2n}$, with
canonical K\"ahler form $\omega = \sum_{i=1}^n dx^i \wedge dy^i$,
where $z^i = x^i + \sqrt{-1} y^i$.  The ($\omega^p/p!$)-grassmannian
is nothing but the grassmannian $G_{\CC}(p,n) \subset G(2p,2n)$ of
complex $p$-planes in $\CC^n$, which is acted transitively by $\SU(n)$
with isotropy $\mathrm{S}(\U(p)\times\U(n-p))$.


If $M$ is also Ricci-flat, so that its holonomy lies in
$\SU(n)$---that is, a Calabi-Yau manifold---then there are in addition
to the K\"ahler calibrations, a circle's worth of special lagrangian
calibrations
\begin{equation*}
\Lambda_\theta = \Re\, e^{i\theta} dz^1 \wedge dz^2 \wedge \cdots
\wedge dz^n~,
\end{equation*}
where $z^i$ are local complex coordinates and $\theta \in S^1$.  Its
calibrated submanifolds are called {\em special lagrangian\/}.  Notice
that the subset of $G(n,2n)$ consisting of all lagrangian planes (with
respect to $\omega$) is not the $\varphi$-grassmannian for any
$\varphi$.  Nevertheless it is fibred over the circle with fibres the
special lagrangian planes relative to $\Lambda_\theta$, for $\theta
\in S^1$.  In other words, every lagrangian plane is special
lagrangian with respect to some $\Lambda_\theta$.  For a lagrangian
submanifold of a Calabi--Yau manifold, the tangent plane at a point
$p$ is a special lagrangian plane relative to $\Lambda_{\theta(p)}$.
Such a manifold is minimal if and only if $\theta$ is constant.  In
other words, a lagrangian submanifold is minimal if and only if it is
special lagrangian.  Notice that $\SU(n)$ acts transitively on the
special lagrangian planes with isotropy $\SO(n)$.

Calibrated geometries also exist for 7- and 8-manifolds of $G_2$ and
$\Spin(7)$ holonomy respectively, as well as for hyperk\"ahler
manifolds.  The exceptional cases are particularly interesting.  On a
7-manifold $M$ of $G_2$ holonomy, there exists a distinguished
parallel 3-form $\varphi$ which is a calibration.  A 3-dimensional
submanifold $N\subset M$ calibrated by $\varphi$ is called {\em
associative\/}.  The grassmannian of associative planes is acted on
transitively by $G_2$ with isotropy $\SO(4)$.  The dual form $\psi =
\star\varphi$ is also a calibration, which calibrates the
4-dimensional {\em coassociative\/} submanifolds of $M$.  An oriented
4-dimensional plane in $\EE^7$ is coassociative if the canonically
oriented normal 3-plane is associative.  Hence the grassmannian of
coassociative planes is also isomorphic to $G_2/\SO(4)$.

Finally, let $\Omega$ denote the parallel self-dual 4-form in an
$8$-dimensional riemannian manifold $M$ of $\Spin(7)$ holonomy.  Then
$\Omega$ is a calibration known as the {\em Cayley\/} calibration.  A
4-dimensional submanifold $N$ calibrated by $\Omega$ is called a {\em
Cayley submanifold\/}.  The grassmannian of Cayley planes of $\EE^8$
is acted on transitively by $\Spin(7)$ with isotropy $\SU(2)\times
\SU(2) \times \SU(2)/\ZZ_2$.  In fact, this is isomorphic to the
grassmannian $G(3,7)$ of oriented 3-planes in $\EE^7$.  This is no
accident, since given any oriented 3-plane in $\EE^7$, there is a
unique Cayley plane in $\EE^8$ which contains it.

\subsection{Calibrations and spinors}

On riemannian spin manifolds with parallel spinors, there is a uniform
construction of the parallel forms in terms of the spinors (see
\cite{Wang}).  This relation makes it easy to compute the comass of
the parallel form and hence to study their grassmannians.

For example, on a manifolds of exceptional holonomy there is a unique 
parallel spinor $\varepsilon$, up to normalisation.  Normalising the
spinor properly one obtains, by squaring,  the calibrations discussed
above \cite{DadokHarvey-spinors}.  Indeed,
\begin{align*}
\varepsilon \otimes \bar\varepsilon &= 1 + \varphi + \psi +
\vol\qquad\text{in the case of $G_2$ holonomy, and}\\
\varepsilon \otimes \bar\varepsilon &= 1 + \Omega + \vol\qquad\text{in
the case of $\Spin(7)$ holonomy.}
\end{align*}
The importance of this construction is that the comass of forms
obtained by squaring spinors is easy to compute in terms of the
spinor.  For example, let $\xi$ be a simple unit 4-vector in $\EE^8$.
Then it follows from the second of the above identities that
$\Omega(\xi) \|\varepsilon\|^2 = \langle \varepsilon, \xi \cdot
\varepsilon\rangle$, where $\|\varepsilon\|^2 = \langle\varepsilon,
\varepsilon\rangle$, $\cdot$ means Clifford action and where
we have used implicitly the isomorphism of the Clifford algebra
$\Cl(8)$ with the exterior algebra $\bigwedge\EE^8$.  By the
Cauchy--Schwarz inequality, it follows that
\begin{equation*}
\Omega(\xi) = \frac{\langle \varepsilon, \xi\cdot
\varepsilon\rangle}{\|\varepsilon\|^2} \leq
\frac{\|\xi \cdot \varepsilon\|}{\|\varepsilon\|}~.
\end{equation*}
Because $\xi$ belongs to $\Spin(8) \subset \Cl(8)$, $\|\xi \cdot
\varepsilon\| = \|\varepsilon\|$, whence $\Omega(\xi) \leq 1$ for all
$\xi$.  In other words, $\Omega$ has unit comass; that is, it is a
calibration.  It follows from this argument that the plane defined by
the 4-vector $\xi$ is calibrated by $\Omega$ if and only if $\xi
\cdot \varepsilon = \varepsilon$.  A slight variant of this argument,
but with spinors of $\Cl(10,1)$, will play a role in our discussion of
intersecting $\M5$-branes.

Two other riemannian holonomy groups also give rise to parallel
spinors: $\SU(n)$ and $\Sp(n)$; and although the analysis is slightly
more involved, the corresponding calibrations can also be built out of
spinors as shown in \cite{Wang}.

\section{The Angle Theorem and the Nance calibration}

An interesting question in the the study of minimal submanifolds is
the following: {\em What are the allowed singularities of a minimally
immersed submanifold?\/} At a point of self-intersection, the tangent
spaces of a minimal $p$-submanifold will form a configuration of
intersecting $p$-planes, and one can ask when such a configuration
will be (locally) volume minimising.  The analogous local question in
the study of branes is {\em What are the allowed supersymmetric
configurations of intersecting branes?\/}  These questions are not
unrelated, since as we will see below supersymmetric brane
configurations are (locally) volume-minimising.

Let us first consider a self-intersection involving only two planes.
The answer to the first question (minimality) is known and goes by the
name of the Angle Theorem,  first conjectured by Morgan
\cite{Morgan-angle} and proven later by the complementary work of
Nance \cite{Nance} and Lawlor \cite{Lawlor} (see also \cite{Harvey}).
The answer to the second question (supersymmetry) is known at least
for the case of $\M5$-branes, from the work of Ohta \& Townsend
\cite{OhtaTownsend}, as we saw in Section 2. In this section we will
briefly discuss the Angle Theorem and the construction of the Nance
calibrations, which are used in the proof of the theorem.  In the next
section we will relate this to the problem of intersecting
$\M5$-branes.

\subsection{The Angle Theorem}

A natural question one can ask is when is the union $\xi \cup \eta$ of
two oriented $p$-planes in $\EE^m$ locally volume-minimising; or
equivalently, when does the two-point set $\{\xi,\eta\}$ belong to the
same $\varphi$-grassmannian, for some $p$-form $\varphi$ of unit
comass.  This question has a simple answer, which is known as the
Angle Theorem.

We will analyse the case of two $p$-planes in $\EE^{2p}$.  The general
case reduces to this one.  Given two $p$-planes $\xi$ and $\eta$,
there is a rotation $R\in\SO(2p)$ which takes one into the other.  We
can always change basis in $\EE^{2p}$ so that $R$ belongs to a chosen
maximal torus.  Hence $R$ is described by $p$ angles $\theta_i$.
These angles are not unique, because any two sets of angles related by
the action of the Weyl group will yield the same configuration of
$p$-planes.  The Weyl group of $\SO(2p)$ is described as follows.
Consider the group of permutations $\sigma$ of the set
$\{-p,\ldots,-1,1,\ldots,p\}$ such that $\sigma(-j) = -\sigma(j)$.
This group is isomorphic to the semidirect product $\fS_p \ltimes
(\ZZ_2)^p$, where the symmetric group $\fS_p$ acts on $(\ZZ_2)^p$
interchanging the factors.  The Weyl group of $\SO(2p)$ is then the
subgroup consisting of even permutations.  Its action on the maximal
torus is given by $(\theta_1,\ldots,\theta_p) \mapsto
(\theta_{\sigma(1)},\ldots,\theta_{\sigma(p)})$ with the convention
that $\theta_{-j} = - \theta_j$.  Up to a Weyl transformation, we can
therefore choose the angles so that $0 \leq \theta_1 \leq \theta_2
\leq \cdots \leq \theta_p$, and such that $\theta_{p-1} \leq
\frac{\pi}{2}$ and $\theta_p \leq \pi - \theta_{p-1}$.

Let $\theta = (\theta_1,\ldots,\theta_p)$ denote a set of such angles
and let $\zeta(\theta)$ denote the unit simple $p$-vector
\begin{equation}\label{eq:zeta}
\zeta(\theta) = (\cos\theta_1 e_1 + \sin\theta_1 e_{p+1}) \wedge
\cdots \wedge (\cos\theta_p e_p + \sin\theta_p e_{2p})~.
\end{equation}

\begin{thm}[Angle Theorem]
Two oriented $p$-planes in $\EE^{2p}$ defined by the two simple unit
$p$-vectors: $\zeta(0) = e_1 \wedge \cdots \wedge e_p$ and
$\zeta(\theta)$, belong to the same $\varphi$-grassmannian if and only
if the following inequality is satisfied:
\begin{equation*}
\theta_p \leq \theta_1 + \theta_2 + \cdots + \theta_{p-1}~.
\end{equation*}
\end{thm}

\subsection{The Nance calibrations}

The `if' part of the Angle Theorem was proven by Nance \cite{Nance} by
explicitly constructing a generalisation of the special lagrangian
calibration tailor-made to calibrate both $\zeta(0)$ and
$\zeta(\theta)$.

Let $u=(u_1,\ldots,u_p)$ be a $p$-tuple of imaginary quaternions of
unit norm, and consider the following $p$-form:
\begin{equation*}
\varphi_u = \Re\, (e^*_1 + u_1 e^*_{p+1}) \wedge \cdots \wedge
(e^*_p + u_p e^*_{2p})~.
\end{equation*}
Let $\zeta(\theta)$ be a $p$-vector as defined above.  If we introduce
the unit quaternions $v_j = \cos\theta_j + \sin\theta_j\,u_j$, then
\begin{equation*}
\varphi_u(\zeta(\theta)) = \Re\, v_1 v_2 \cdots v_p \leq | v_1 v_2
\cdots v_p| = |v_1| |v_2| \cdots |v_p| = 1~.
\end{equation*}
This does not yet show that $\varphi_u$ has unit comass, because not
every unit simple $p$-vector is of the form $\zeta(\theta)$ in this
basis.  However it follows from Morgan's Torus Lemma (see
\cite{Harvey}) that $\varphi_u$ attains its maximum on a subset of the
grassmannian which always contains at least one such $p$-vector.
Hence modulo this lemma, we have shown that $\varphi_u$ has unit
comass.  Furthermore $\zeta(\theta)$ is calibrated by $\varphi_u$
precisely when $v_1 v_2 \cdots v_p =1$.  Therefore to construct the
{\em Nance calibration\/} $\varphi_u$ it is necessary to find $p$ unit
quaternions $v_j$ satisfying $v_1 v_2 \cdots v_p=1$ with $\Re\,v_j =
\cos\theta_j$.  The $u_j$ are then defined by $u_j = \Im\,v_j /
|\Im\,v_j|$.

The condition $v_1 v_2 \cdots v_p=1$ is automatically satisfied if we
introduce $p$ unit quaternions $w_j$ such that
\begin{equation*}
v_j = w_j \Bar w_{j+1}\qquad\text{with the convention that $w_{p+1} =
w_1$.}
\end{equation*}
We therefore need to find $p$ unit quaternions $w_j$ such that
$\Re\, w_j \Bar w_{j+1} = \cos\theta_j$.  To make matters easier, let
us choose the $w_j$ to be imaginary quaternions.  Then because they
have unit norm, they define points on the unit 2-sphere $S^2 \subset
\Im\,\HH$.  The condition that $\Re\, w_j \Bar w_{j+1} = \cos\theta_j$
is equivalent to the spherical distance (along great circles)
$d(w_j,w_{j+1})$ being equal to $\theta_j$.  Hence finding the $w_j$
is equivalent to finding a spherical $p$-gon with sides $\theta$.
Because of the conditions on the $\theta_j$, this is only possible if
$\theta_p \leq \theta_1 + \cdots + \theta_{p-1}$.

\section{Intersecting $\M5$-branes at angles}

We are now in a position to re-interpret the results of Ohta \&
Townsend \cite{OhtaTownsend} in terms of calibrations, and to rederive
them using the relationship between spinors and calibrations.

\subsection{Special lagrangian calibrations and the condition
\eqref{eq:msusy}}

The initial observation which led us to the work reported here is the
fact that the condition for supersymmetry \eqref{eq:msusy} is
precisely the case where the inequality in the Angle Theorem is
saturated.  As we now show, this implies that the Nance calibration is
special lagrangian.

When the angle inequality is saturated, the vertices $w_j$ of the
spherical $p$-gon in the construction of the Nance calibration all lie
on the same great circle.  Thinking of this great circle as the
equator, then the $u_j$ are all equal to the same unit imaginary
quaternion $q$: the north pole.  We can always orient the 
sphere so that $q=i$, which brings the Nance calibration to the
special lagrangian form:
\begin{equation}\label{eq:slagform}
\varphi_u = \Re\, (e^*_1 + i e^*_{p+1}) \wedge \cdots \wedge (e^*_p +
i e^*_{2p})~.
\end{equation}
Therefore the two planes $\xi$ and $\eta$ are special lagrangian
planes with respect to the same special lagrangian calibration.  In
particular, this means that they are related by an element of
$\SU(p)$.

Applying this to the case of two intersecting $\M5$-branes, we see
that the configuration is supersymmetric precisely when the two branes
are {\em simultaneously calibrated by the same special lagrangian
5-form.\/} In particular, they are related not by a general $\SO(10)$
rotation but in fact by an $\SU(5)$ rotation. This is in agreement
with \cite{BerkoozDouglasLeigh} and provides a geometric underpinning
for their result.

\subsection{Spinors and the generalisation to more than two branes}

In fact, these results can be understood at a more conceptual level in
terms of spinors.  This will have the added benefit of allowing us to
generalise this to any number of intersecting branes.  Let
$\varepsilon$ again denote a real (Majorana) spinor of $\Cl(10,1)$.
Squaring the spinor we obtain on the right-hand side a 1-form, a
2-form and a 5-form:
\begin{equation}\label{eq:square}
\varepsilon \otimes \bar\varepsilon = \Omega^{(1)} + \Omega^{(2)} +
\Omega^{(5)}~,
\end{equation}
where by $\bar\varepsilon \equiv  -(e_0 \cdot \varepsilon)^t$ we mean
the Majorana conjugate.  In this expression $\Omega^{(p)}$ is a
$p$-form in $\MM^{10,1}$.  Under the orthogonal decomposition
$\MM^{10,1} = \EE^{10} \oplus \RR e_0$, the 5-form $\Omega^{(5)}$
breaks up as
\begin{equation}\label{eq:bilinear}
\Omega^{(5)} = e_0^* \wedge \Theta^{(4)} + \Theta^{(5)}~,
\end{equation}
where $\Theta^{(4)}$ and $\Theta^{(5)}$ are a 4- and a 5-form on
$\EE^{10}$, respectively.  Now let $\xi$ be an oriented 5-plane in
$\EE^{10}$ and consider the bilinear $\bar\varepsilon \xi \cdot
\varepsilon$.  Using \eqref{eq:bilinear} and the definition of the
Majorana conjugate, one can rewrite this as
\begin{equation*}
\langle \varepsilon, (e_0\wedge \xi )\cdot \varepsilon\rangle =
\Theta^{(5)}(\xi) \tr \1 = 32\,\Theta^{(5)}(\xi)~,
\end{equation*}
where we have introduced the $\Spin(10)$-invariant inner product
$\langle -,-\rangle$ defined by $\langle \chi,\varepsilon\rangle =
\chi^t \varepsilon$.  By the Cauchy--Schwarz inequality for this inner
product, we find that
\begin{equation*}
\Theta^{(5)}(\xi) \leq \tfrac{1}{32} \|\varepsilon\|
\|(e_0\wedge\xi)\cdot\varepsilon\|~.
\end{equation*}
Because $\xi$ is a unit simple 5-vector, $\|(e_0
\wedge\xi)\cdot\varepsilon\| = \|\varepsilon\|$, whence
\begin{equation*}
\Theta^{(5)}(\xi) \leq \tfrac{1}{32} \|\varepsilon\|^2.
\end{equation*}
In other words, the comass of $\Theta^{(5)}$ is given by
$\frac{1}{32}\|\varepsilon\|^2$, and a 5-plane $\xi$ is calibrated by
$\Theta^{(5)}$ if and only if $(e_0\wedge\xi) \cdot \varepsilon =
\varepsilon$, which is precisely the condition \eqref{eq:fundsusy} for
supersymmetry of the $\M5$-brane.  Therefore if $\xi$ and $\eta$ are
two 5-planes tangent to $\M5$-branes, so that they satisfy
\eqref{eq:fundsusy}, they are both calibrated by the same 5-form
$\Theta^{(5)}$.

{\em What about more than two intersecting branes?\/} A configuration
of $m$ intersecting 5-planes $\cup_i \xi_{(i)}$ is supersymmetric if
and only if there exists a nonzero spinor $\varepsilon$ such that 
\begin{equation*}
(e_0 \wedge \xi_{(i)})\cdot \varepsilon = \varepsilon\qquad\text{for
all $i$.}
\end{equation*}
From the above discussion, this means that they are simultaneously
calibrated by the same calibration $\Theta^{(5)}$, so that they belong
to the same $\Theta^{(5)}$-grassmannian.  Just as for minimality (see, 
for example, \cite{Lawlor-kplanes}), it is not known whether it is
sufficient to demand that pairwise intersections be supersymmetric.

Although we have shown above that any two intersecting planes
resulting in a supersymmetric configuration are calibrated by the
special lagrangian 5-form $\varphi_u$ in \eqref{eq:slagform}, and that
they are also calibrated by the same 5-form $\Theta^{(5)}$ in
\eqref{eq:bilinear}, it does not necessarily follow that
$\Theta^{(5)}$ is a special lagrangian calibration.  It is possible
for a plane or planes to be simultaneously calibrated by two different
forms.  In order to identify $\Theta^{(5)}$ one has to work a little
harder.\footnote{The following discussion owes a great deal to Robert
Bryant.}

The nature of $\Theta^{(5)}$ depends on the isotropy group of the
spinor $\varepsilon$.  A nonzero Majorana spinor $\varepsilon$ of
$\Spin(10,1)$ can have two possible isotropy groups: either $\SU(5)
\subset \Spin(10)$ acting trivially on the time-like direction, or a
30-dimensional non-semisimple Lie group $G \cong \Spin(7) \ltimes
\RR^9$, where $\Spin(7)$ acts on $\RR^9$ as $\RR^8 \oplus \RR$, the
first factor being a spinor and the second factor being the trivial
representation.  In the former case, the 5-form $\Theta^{(5)}$ is
$\SU(5)$-invariant and is therefore a special lagrangian calibration,
whereas in the latter case, the 5-form is of the form $v^* \wedge
\Omega$ where $\Omega$ is a Cayley calibration in an eight-dimensional
subspace $V \subset \EE^{10}$ and $v \in V^\perp$ is a fixed vector
perpendicular to $V$.  In particular this means that in the Cayley
case, the tangent planes to the branes are actually $\Spin(7)$
related, so that only four of the five angles are nonzero: they
intersect generically on a one-dimensional subspace, not on a
point.  In summary, for a special lagrangian $\Theta^{(5)}$, a
supersymmetric configuration of $m$ intersecting $\M5$-branes at
angles in general position preserves $\frac{1}{32}$ of the
supersymmetry, whereas for the $G$-invariant $\Theta^{(5)}$, there is
twice as much supersymmetry preserved if $m{=}2$, and again
$\frac{1}{32}$ if $m>2$.

Geometrically, we can interpret this result by saying that a
supersymmetric configuration of any $m$ 5-planes calibrated by the
special lagrangian $\Theta^{(5)}$ can be interpreted as the
self-intersection of a special lagrangian submanifold of $\EE^{10}$,
whereas when $\Theta^{(5)}$ is $G$-invariant, it is to be interpreted
as the self-intersection of a Cayley submanifold of a fixed $\EE^8
\subset \EE^{10}$.


\subsection{A brief look at $\M2$-branes}

A similar analysis holds for $\M2$-branes. Let $\zeta$ and $\chi$ now
denote the tangent 2-planes to a pair of static $\M2$-branes.  This
configuration will be supersymmetric provided that the analogue of
\eqref{eq:fundsusy} holds:
\begin{equation}\label{eq:susym2}
(e_0 \wedge \zeta) \cdot \varepsilon =
\varepsilon\qquad\text{and}\qquad (e_0 \wedge \chi) \cdot \varepsilon
= \varepsilon~,
\end{equation}
for some spinor $\varepsilon$.  Squaring the spinor yields
\eqref{eq:square}, and we choose to decompose $\Omega^{(2)}$ as
\begin{equation*}
\Omega^{(2)} = e_0^* \wedge \Theta^{(1)} + \Theta^{(2)}~,
\end{equation*}
where $\Theta^{(1)}$ and $\Theta^{(2)}$ are a 1- and a 2-form on
$\EE^{10}$ respectively.  Then the same argument as for the
$\M5$-branes, shows that the pair of 2-planes $\zeta$ and $\chi$ are
simultaneously calibrated by $\Theta^{(2)}$.  Moreover, any
configuration of $m$ 2-planes simultaneously calibrated by
$\Theta^{(2)}$ will also be supersymmetric for any $m$.  Again the
nature of $\Theta^{(2)}$ depends on the isotropy group of the spinor.
If the spinor is $\SU(5)$-invariant then $\Theta^{(2)}$ is a K\"ahler
form, so that the configuration would correspond to the
self-intersections of a complex curve in $\EE^{10}$.  Alternatively,
if the spinor is $G$-invariant, then $\Theta^{(2)}$ is the volume form
of the perpendicular subspace to the $\EE^8\subset\EE^{10}$ singled
out by the $\Spin(7) \subset \SO(8)$ subgroup of $\SO(10)$.  In this
case, the $\M2$-branes are all coincident and the configuration
preserves $\half$ of the supersymmetry.

\section{The many faces of susy}

The configurations described above are those in which the $\M$-branes
are in general position.  There is a rich variety of special
configurations which preserve more of the supersymmetry.  For a pair
of $\M5$-branes, these configurations have been classified by Ohta \&
Townsend \cite{OhtaTownsend}.  A group-theoretical interpretation of
their work will be reported on elsewhere \cite{InProgress}.  For more
than two branes, the problem is still open; although some of these
configurations, including some which mix $\M2$- and $\M5$-branes, have
been studied in a similar context to the one in this paper in
\cite{GibbonsPapadopoulos} and from the point of view of the
worldvolume theory in \cite{GLW}.  Nevertheless there is still much to
be done to reach some sort of classification.  It might help to
compare the status of this problem with the analogous problem in the
theory of minimal surfaces.

The study of the faces of the grassmannian of oriented $p$-planes is
an important problem in the theory of minimal surfaces, and one which
has not been solved except in the simplest of cases.  The faces of the
grassmannian of lines and of 2-planes are of course classical: the
calibrated submanifolds correspond to real and complex lines.  The
first non-trivial example is thus the grassmannian of oriented
3-planes in $\EE^n$ for $n\geq 6$, which was studied in
\cite{DadokHarvey-R6,HarveyMorgan-6,Morgan-ext}.  To this date, the
faces of the grassmannian of oriented $p$-planes in $\EE^n$ have been
worked out fully only for the following other values of $(p,n)$:
$(3,7)$ (which is equivalent to $(4,7)$) in \cite{HarveyMorgan-7} ,
and $(4,8)$ in \cite{DadokHarveyMorgan}.

Fortunately, we are not interested in just any face of the
grassmannian, but only in the supersymmetric faces; that is, those
corresponding to calibrations which are associated to spinors in the
way described above.  This problem appears much more tractable.  Many
of these faces are known: associative and coassociative planes in
$\EE^7$, Cayley planes in $\EE^8$ and special lagrangian planes in
$\EE^{2n}$; but a complete classification, at least in low dimension,
say $d\leq10$, is lacking.  Also one should not just treat euclidean
or static branes, but arbitrary branes on lorentzian signature (see
\cite{Mealy}).  We hope to report on these generalisations on future
work \cite{InProgress}.

\section*{Acknowledgements}

It is a pleasure to thank Frank Morgan and especially Robert Bryant
for helpful correspondence, Sonia Stanciu for useful discussions and
her critical comments on a previous version of this paper, and Gary
Lawlor for sending us a copy of \cite{Lawlor-kplanes}. BSA is
supported by a PPARC Postdoctoral Fellowship, JMF by an EPSRC PDRA and
BS by an EPSRC Advanced Fellowship, and we would like to extend our
thanks to the relevant research councils for their support.

%
\providecommand{\bysame}{\leavevmode\hbox to3em{\hrulefill}\thinspace}

\end{document}